# Existence of Unexcited and Excited Biexcitons in Molecular Crystals

**Mohamed Assad Abdel-Raouf, Physics** *Department, E-mail:* [assad@uaeu.ac.ae](mailto:assad@uaeu.ac.ae) *,*

*U.A.E. University, Al-Ain, P.O. Box: 17555, U.A.E.*

## Abstract

The theory of four-body systems is revisited. It is illustrated that the theory provides a rigorous proof for the formation of ground state (unexcited) biexcitons in molecular crystals. The generalization of the theory predicts the possible existence of excited biexcitons in nature. In order to test the validity of the extended theory on computational level, a very elaborate computer code is constructed for the treatment of arbitrary four-body system with arbitrary electron/hole mass ratio ($\sigma$) and arbitrary exciton - exciton interaction ($V_{XX}$). The results, for $\sigma = 1$ and $V_{XX}$ is a pure Coulomb force, show that the test is successful. Realization of these conclusions on experimental level should open the door for wide industrial applications.

--------------------------------------------------------------------------------

# 1. INTRODUCTION

In solid state physics and material science, the pioneering works of Frenkel [1], Peierls [2] and Wannier [3] have established the fact that a crystal in the ground state, i.e. with empty conduction band and unexcited valence electrons, may absorb a photon and one of its Fermi electrons could jump to the conduction band leaving a positive hole behind. The electron should feel the attractive force of the hole in form of a pure Coulomb or partially screened force. Both quasi particles perform together a quasi ground state called exciton which is the main reason for a semiconductor (e.g. gallium arsenide and aluminum gallium arsenide) to act as an electric device. An exciton may collapse, (when the quasi particles are recombined together), producing a gamma photon and unexcited crystal. (For this reason excitons are known as light sources in semiconductors). This amazing phenomenon has gained much interest over the years [4]. A drastic development in the theory of excitons was achieved when Lampert [5] and Moskalenko [6] predicted the existence of a ground- state composed of two excitons. They called the new four-body structure "biexciton". Decisive confirmation of the existence of biexcitons (which are also referred to as "'excitonic molecules") was given by Haynes [7]. The simplest (ideal) kind of these molecules are the so called Wannier's biexcitons. These are composed of two identical quasi particles, each has an effective mass m and an effective charge – Ze, in addition to other two identical quasi particles, each of effective mass M and an effective charge Ze, where e is the electronic charge and $Z = 1/\varepsilon_0$. The parameter $\varepsilon_0$ stands for the dielectric constant of the crystal. It is interesting to mention here that Wannier's excitons are distinguished from other excitons (e.g. the Frenkel ones) by their relatively large electron-hole distances and the domination of the long range Coulomb forces. Therefore, depending on the main interaction force between the quasi particles, an analogy is made in the literature between biexcitons and other four-body molecules appearing in atomic, molecular and high energy physics. Earlier theoretical calculations of the ground state energies of various excitonic molecules were performed by many authors, (for a review see Abdel-Raouf [8]). The question, however, about the existence of excitonic molecules with arbitrary value of $\sigma$ was settled down by the theorem of four-body of systems [9]. In one of the corollaries of the theorem it was shown that the existence of molecular structures with $\sigma = 0$ and $\sigma = 1$ in the k th states is sufficient for the existence of the $k^{th}$ states of biexcitons of arbitrary $\sigma$. On the other hand, the last decade has shown tremendous interest in the experimental and theoretical investigations [10] of the ground states of biexcitons created in large varieties of semiconductors (e.g. GaAs/AlGaAs). In particular, the last few years, have seen a new fundamental development in the theory of biexcitons, namely the experimental evidences which suggest the existence of excited biexcitons [11]. These can be interpreted as bound states between excited and unexcited excitons or created when biexcitons absorb photons.

The aim of the present work is two fold: (1) to recall the theory of four-body systems and the corollaries mentioned above, (2) to test the possible existence of excited biexcitons using our new code. Apart from the present section, the paper contains other three sections. Section 2 involves a brief account on the theory-of four-body systems. Section 3 is devoted to the variational treatment of four-body molecules. In section 4 the results and discussion of our calculations are presented.

## 2. THEORY OF FOUR-BODY SYSTEMS

Following Abdel-Raouf [9], a theorem was proved which guarantees the existence of any four-body molecule under the assumption that the mass and charge symmetry of the system is fulfilled. Particularly, two principle conclusions were drawn out of the theorem:

(i) The binding energies of the hydrogen and positronium molecules (denoted as $W_1(0)$ and $W_1(1)$, respectively) are lower and upper bounds, respectively, on the binding energies ($W_1(\sigma)$) of all four-body molecules with $\sigma$ falling in the region [0,1].

(ii) The binding energy $W1(\sigma)$ is a concave function of $\sigma$ lying inside the triangle $(0, W_1(1))$, $(0, W_1(0))$ and $(1, W_1(1))$, (see Fig (1)).

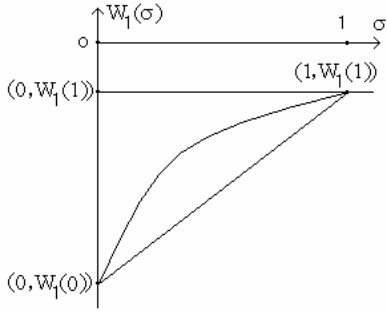

**Fig. 1: Behavior of $W_1(\sigma)$ with $\sigma$.**

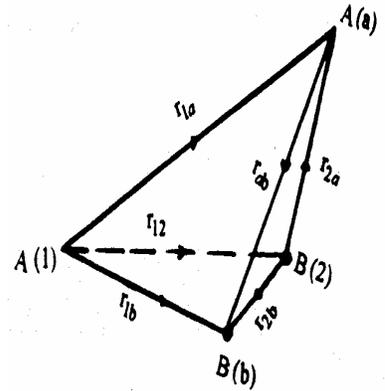

Figure 2. Relative coordinates of the four-body molecules.

Consider now a quantum mechanical system ABBA composed of two pairs of particles and antiparticles (Fig. 2), where the subclusters AB and ABA or BAB exist in atomic and ionic forms respectively. Let m and M be the masses of the two particles A and B respectively. Assuming that the molecular bound state of the system -if it exists- is governed by the two-body Coulomb force between each pairs of particles, the total Hamiltonian of the system can be expressed as:

$$H = -\frac{\hbar^2}{2m}(\nabla_1^2 + \nabla_a^2) - \frac{\hbar^2}{2M}(\nabla_2^2 + \nabla_b^2)$$
$$+ Z^2 e^2 \left(\frac{1}{r_{12}} + \frac{1}{r_{ab}} - \frac{1}{r_{1a}} - \frac{1}{r_{2a}} - \frac{1}{r_{1b}} - \frac{1}{r_{2b}}\right) \quad (1)$$

$$= -\frac{\hbar^2}{2mM}[M(\nabla_1^2 + \nabla_a^2) + m(\nabla_2^2 + \nabla_b^2)]$$
$$+ Z^2 e^2 \left(\frac{1}{r_{12}} + \frac{1}{r_{ab}} - \frac{1}{r_{1a}} - \frac{1}{r_{2a}} - \frac{1}{r_{1b}} - \frac{1}{r_{2b}}\right). \quad (2)$$

Defining the reduced mass μ and the mass ratio σ as;

$$\frac{1}{\mu} = \frac{1}{m} + \frac{1}{M} = \frac{M+m}{mM}, \quad (3)$$

$$\sigma = \frac{m}{M}. \quad (4)$$

Then, the Hamiltonian can be written in the form;

$$H = -\frac{\hbar^2}{2\mu(M+m)}[M(\nabla_1^2 + \nabla_a^2) + m(\nabla_2^2 + \nabla_b^2)] + Z^2 e^2 \left(\frac{1}{r_{12}} + \frac{1}{r_{ab}} - \frac{1}{r_{1a}} - \frac{1}{r_{2a}} - \frac{1}{r_{1b}} - \frac{1}{r_{2b}}\right)$$

$$= -\frac{\hbar^2}{2\mu}[\frac{1}{1+\sigma}(\nabla_1^2 + \nabla_a^2) + \frac{\sigma}{1+\sigma}(\nabla_2^2 + \nabla_b^2)] + Z^2 e^2 \left(\frac{1}{r_{12}} + \frac{1}{r_{ab}} - \frac{1}{r_{1a}} - \frac{1}{r_{2a}} - \frac{1}{r_{1b}} - \frac{1}{r_{2b}}\right). \quad (5)$$

Now, considering the Rydberg units where;

$$\left.\begin{array}{l}\dfrac{\mu Z^4 e^4}{2\hbar^2} \text{ is the energy unit} \\ \dfrac{\hbar^2}{\mu Z^2 e^2} \text{ is the length unit}\end{array}\right\} \quad (6)$$

Taking $\dfrac{\hbar^2}{\mu Z^2 e^2}$ to be equal to unity, then the Hamiltonian could finally be given by,

$$H = \left\{\dfrac{-1}{1+\sigma}[\nabla_1^2 + \nabla_a^2 + \sigma(\nabla_2^2 + \nabla_b^2)]\right\}_T$$
$$+ \left\{\dfrac{2}{r_{12}} + \dfrac{2}{r_{ab}} - \dfrac{2}{r_{1a}} - \dfrac{2}{r_{2a}} - \dfrac{2}{r_{1b}} - \dfrac{2}{r_{2b}}\right\}_V \,, \quad (7)$$

where V is its total potential energy, T is the total kinetic energy operator of the system, while

$$\nabla_i^2 = \sum_{j \neq i}\left(\dfrac{\partial^2}{\partial r_{ij}^2} + \dfrac{2}{r_{ij}}\dfrac{\partial}{\partial r_{ij}}\right) + 2\sum_{j \neq i}\sum_{k \neq i}\cos\theta_{ij,ik}\dfrac{\partial^2}{\partial r_{ij}\partial r_{ik}} \quad (8)$$

and $$\cos\theta_{ij,ik} = \dfrac{r_{ij}^2 + r_{ik}^2 - r_{jk}^2}{2\, r_{ij}\, r_{ik}} \quad . \quad (9)$$

Now, if $\{|\psi_k\rangle\}$ is the set of exact wavefunctions of ABBA, such that:

$$\psi_k = \psi_k(r_1, r_2, \ldots\ldots\ldots) \quad (10)$$

$$\langle\psi_k|\psi_{k'}\rangle = \int \psi_k^* \psi_{k'} \, d\tau \,, \quad (11)$$

where $d\tau$ is the volume element, therefore, the bound-states of the system are identical with the negative spectrum of the Hamiltonian within the space $\{|\psi_k\rangle\}$, i.e., they are the eigenvalues of the Schrödinger equation;

$$H|\psi_k\rangle = E_k|\psi_k\rangle \ , \tag{12}$$

and can be determined by;

$$E_k = \langle\psi_k|H|\psi_k\rangle \ , \tag{13}$$

such that $\quad E_k \leq E_{k+1}, \quad$ for all $k \geq 1$ . $\tag{14}$

Obviously, if $E_k \geq 0$ for all k's, then the total Hamiltonian H does not possess any negative spectrum and the quantum mechanical system can not form a bound-state, in other words, the molecule consisting of the four particles A, B, B, A simply can not exist.

Starting from the total exact wavefunction (10), the four-body theorem leads to the following conclusions:

(1) If the molecule ABBA exists at $\sigma = 0$ and $\sigma = 1$, then all molecules with $\sigma$ lying between 0 and $\infty$ must exist.

(2) If $W_k = E_k - 2E_k^{AB}$, where $E_k^{AB}$ is the k-th state of the pair AB, the following two conclusions are true:

(a) If the k-th state of ABBA exists at $\sigma = 0$ and $\sigma\sigma = 1$, the k-th state of ABBA exists at all $0 \leq \sigma \leq 1$.

(b) $W_k(0) \leq W_k(\sigma) \leq W_k(1)$ $\tag{15}$

Conclusion (1) means that the existence of the hydrogen molecule and Wannier's biexciton (at $\sigma = 1$) are sufficient for the existence of all intermediate biexcitons. (where "existence" here indicates formation in the ground state).

Conclusion (2), on the other hand, emphasizes that any experimental or computational proof for the existence Wannier's biexciton in one or all excited states implies the existence of all biexcitons with intermediate mass ratio. It is important to mention that the four body theory guaranties and predicts in principle the possible formation of all biexcitons with intermediate mass ratio σ. To provide a computational proof for the existence of excited Wannier's biexcitons at σ = 1, is the main goal of the present work. In the next section we discuss the problem shortly.

## 3. VARIATIONAL TREATMENT

### 3.1. Rayleigh-Ritz Variational Method

In RRVM a trial wavefunction is expanded by

$$\psi_t^{(n)} = \sum_{k}^{n} a_k |\psi_{tk}\rangle, \quad (16)$$

where n is the dimension of $|\psi_{tk}^{(n)}\rangle$, and

$$\langle \psi_{tk}^{(n)} | \psi_{tk'}^{(n)} \rangle = \delta_{kk'}, \quad \text{for } k, k' = 1, 2, \ldots, n, \quad (17)$$

where $\delta_{kk'}$ is the Kroneker-delta. The linear parameters at (15) are subjected to the variational principle

$$\partial \langle \psi_t^{(n)} | H - E | \psi_t^{(n)} \rangle = 0 \quad (18)$$

All $|\psi_{tk}^{(n)}\rangle$'s are, due to RRVM, generated from one basis set of functions $\{|\chi_i\rangle\} \subset D_H$ where $D_H$ is the H-domain, i.e.,

$$\left|\psi_{tk}^{(n)}\right\rangle = \sum_{i=1}^{n} c_{ik}\left|\chi_i\right\rangle . \tag{19}$$

Consequently, eq. (16) can be written for each k as the system of secular equations:

$$\sum_{j=1}^{n} c_{jk}[\langle\chi_i|H|\chi_j\rangle - E_{nk}\langle\chi_i|\chi_j\rangle] = 0, \quad i = 1, 2, \ldots, n \tag{20}$$

which is meaningful if and only if the determinant $\Delta_{nk}$ vanishes, i.e.

$$\Delta_{nk} = \det(H_{ij} - E_{nk} S_{ij}) = 0, \tag{20a}$$

where

$$H_{ij} = \langle\chi_i|H|\chi_j\rangle \quad \text{and} \quad S_{ij} = \langle\chi_i|\chi_j\rangle. \tag{20b}$$

The eigenvalues obtained by (20) are ordered such that;

$E_{n1} \leq E_{n2} \leq \ldots \leq E_{nn}$.

Also, the important relation between $E_{n1}$ and the first (lowest) exact energy level of the system $E_1$ was proved, namely $E_1 \leq E_{n1}$, for $n > 0$, i.e., for any choice of the components $\left|\psi_{tk}^{(n)}\right\rangle$, the first variational energy is an upper bound to the exact one. Moreover, it was shown that if the condition (20) is fulfilled, one gets

$E_k \leq E_{nk}$, for $k = 1, 2, \ldots n$ .

It has been also demonstrated that, if the number of components of the trial; wavefunction is enlarged by exactly one component such that

$$\left\langle \psi_{tk}^{(n+1)} \middle| \psi_{tk'}^{(n+1)} \right\rangle = \delta_{kk'}, \quad \text{for k, k'} = 1, 2, \ldots, n+1.$$

Thus, the following successive relations are always valid:

$$\begin{aligned}
&E_{n+11} \leq E_{n1} \leq E_{n+12} \leq \ldots \leq E_{n+1n} \leq E_{nn} \leq E_{n+1n+1}, \\
&E_{n+1k} \leq E_{n+1k'} \text{ for } k \leq k', \\
&E_{n+1k+1} \leq E_{nk'} \text{ for } k > k', \\
&E_k \leq E_{n+1k} \leq E_{nk} \leq E_{n-1k} \leq \ldots \leq E_{1k}.
\end{aligned} \quad (21)$$

The main drawback of relations (21), however, lies in the fact that if H does not possess a negative spectrum in $\{|\psi_{tk}^{(n)}\rangle\}$, we are unable to perform any conclusive judgment about the existence or nonexistence of the quantum mechanical system considered. This can be simply attributed to the fact that if all $E_{nk}$'s lie in the continuous spectrum of H, there may exist at least one bound state of ABBA, say $E_1$, but $\{|\psi_{tk}^{(n)}\rangle\}$ ceases to see it because all values $\langle \psi_{tk}^{(n)}|H|\psi_{tk}^{(n)}\rangle$, including the lowest, are upper bounds of $E_1$. This is also true if one or n of $E_{nk}$'s, $n < k$, lie in the negative spectrum of H, in this case, we will be unable to make any decisive statement about all states higher than $E_k$. On the other hand, the existence of any negative $E_{nk}$ ensures the existence of corresponding bound state of ABBA, and for this reason we are led to say that

Rayleigh-Ritz' approximate solution of (10) is sufficient but not necessarily decisive for the existence of the four-body molecule ABBA.

### 3.2. Wavefunction and Hamiltonian

Our trial wavefunction should be developed following eq. (19) from a basis set, the j th component of this set has the form:

$$|\chi_j\rangle = s_1^{m_j} s_2^{n_j} e^{-\alpha_j(s_1+s_2)} t_1^{k_j} t_2^{\ell_j}$$

$$\cosh[\beta_j(t_1-t_2)] u^{p_j} v^{q_j} e^{-\gamma_j v} \quad . \tag{22}$$

where
$$s_i = (r_{ia} + r_{ib})/r_{ab} \quad ; \quad i = 1, 2$$
$$t_i = (r_{ia} - r_{ib})/r_{ab} \quad ; \quad i = 1, 2$$
$$s_a = (r_{1a} + r_{2a})/r_{12} \quad , \quad s_b = (r_{1b} + r_{2b})/r_{12} ,$$
$$t_a = (r_{1a} - r_{2a})/r_{12} \quad , \quad t_b = (r_{1b} - r_{2b})/r_{12} ,$$
$$u = r_{12}/r_{ab} \quad , \quad v = r_{ab}.$$

The Hamiltonian of the four-body biexcitons is defined by eqs. (7) - (9). Applying the kinetic energy operators given at eq. (8) to the j th component $\chi_j$ of the wavefunction, we obtain rather complicated relations which could not be given here due to space restriction. Finally, applying the exact four-body potential energy operator eq. (7), to the j th component of the wavefunction, we find

$$V\chi_j = \frac{2}{v}[(\frac{1}{u}+1) - \frac{4s_1}{s_1^2 - t_1^2} - \frac{4s_2}{s_2^2 - t_2^2}]\chi_j$$

## 4. RESULTS AND DISCUSSIONS

The wavefunction and the Hamiltonian of the excitonic molecule were employed for the calculation of the matrix elements (20b). After the development of the Hamiltonian and overlap matrices, Raleigh-Ritz variational method was used for calculating the binding energies and wavefunctions of the system in the ground state. It is interesting to know that all programs and subprograms were created by us and represent our code for solving the four-body problem in the ground states. The biexcitons considered here are composed of two unexcited excitons, i.e. with zero angular momenta and zero spins. The code was then generalized to deal with unexcited excitons (zero angular momentum and zero spin) interacting with excited excitons (with higher angular momentum and different spins). The coupling of these excitons is decisive for the existence of the excited biexcitons. In the first run of the extended code we considered two excitons of zero spins (both are singlet), however, one of them lies in the s state (zero angular momentum) and the other exists in the p state (total angular momentum = 1). The number of the components of the basis set employed for the treatment of unexcited and excited biexcitons was steadily increased up to 30 components. (Eight functions more that what we used in the last treatment of positronium molecules [12]. The resultant binding energy of the unexcited system (-4.93 eV), (i.e. the total energy = -18.53 eV) has been slightly improved in comparison with the last one (-4.710 eV). The same number of components of the basis set has been employed for the evaluation of the binding energy of the excited biexcitons described previously. The binding energy of this state was found to be -1.271 eV. This means that the molecule is stable against dissociation into an exciton in the ground state and an excited exciton in the p state. This result is of fundamental consequences, since it confirms the possible existence of excited biexcitons in molecular crystals, semiconductors, superlattices, etc. In spite of this conclusion, however, much work should be done in order to improve the resultant binding energy and to prove the existence of these states for arbitrary values of $\sigma$.

**Figures Captions:**

**Fig. 1: Behavior of $W_1(\sigma)$ with $\sigma$.**

**Fig. 2: Relative Coordinates of the four-body Molecule**